# CHEMICAL ANALYSES OF SILICON AEROGEL SAMPLES

I. van der Werf[a], F. Palmisano[a], R. De Leo[b], S. Marrone[b]

*a) Dipartimento di Chimica, Università di Bari, Bari, Italy.*
*b) Dipartimento di Fisica, Università di Bari and INFN Sezione di Bari, Bari, Italy.*

**Abstract**

After five years of operating, two Aerogel counters: A1 and A2, taking data in Hall A at Jefferson Lab, suffered a loss of performance. In this note possible causes of degradation have been studied. In particular, various chemical and physical analyses have been carried out on several Aerogel tiles and on adhesive tape in order to reveal the presence of contaminants.

April 9[th] 2008

# TABLE of CONTENTS

# 1. Introduction

In the year '00 two Cherenkov counters with radiators of silica Aerogel with different refractive indices, denoted A1 and A2, were installed in the focal plane equipment of the hadron spectrometer in Hall A for the identification of pions, protons and kaons. In the following years several Hall A experiments have successfully employed these counters with performance very near to that of the first installation. In '04, during the first runs of the experiment E94-107, the photoelectron (pe) distributions both in A1 and in A2 were found with an average pe number noticeably reduced with respect to the original one. This degradation was due to the outer layers of Aerogel, which at a visual inspection looked opaque and of yellow colour. The contamination causes were initially attributed to the diffusing material (Millipore paper) and to the adhesive tape covering the counter internal surfaces found broken and crumbled in several parts. Another factor of contamination was identified in the air forced to circulate in the counters to avoid the accumulation of Helium gas released from the cryogenic systems installed in Hall A. In order to keep the counters fully performing for the high precision hypernuclear experiment E94-107 in '05, the degraded aerogel layers and the diffusing material were immediately replaced.

In order to determine the exact causes of the contaminations, many chemical and physical analyses have been carried out on several Aerogel samples and on adhesive tape. Table 1 reports the samples of aerogel (used – configuration '04 - and unused – configuration '05 -) and adhesive tape analyzed together with the techniques used to perform such study.

Table 1: Analyses performed on different samples of silicon aerogel and adhesive tape.

| SAMPLE | SPME-GC/MS | CONTACT ANGLE | GC (EXTRACT) | UV-VIS SPECTROSCOPY |
|---|---|---|---|---|
| SP15 ('04 configuration) | + | - | - | + |
| SP30 ('05 configuration) | + | - | - | - |
| SP50 ('04 configuration) | + | + | + | + |
| SP50 ('05 configuration) | + | + | + | + |
| Adhesive tape | + | - | - | - |

# 2. Solid Phase Micro Extraction-Gas Chromatography/Mass Spectrometry

*2.1 Experimental: Samples and Analytical Procedures*

Samples of Matsushita silicon Aerogel (used – configuration '04 - and unused – configuration '05 -) with different refractive indices: SP15 ($n = 1.015$), SP30 ($n = 1.030$), SP50 ($n = 1.050$) were investigated by Solid Phase Micro Extraction-Gas Chromatography/Mass Spectrometry (SPME-GC/MS). Small amounts of these aerogels and of the adhesive tape (0.5 - 1.7 grams) were introduced into 40 mL glass vials and heated at ~ 85° C for twenty minutes. Then a SPME fibre (divinylbenzene-carboxen-polydimethylsiloxane) by Supelco [Sup] was exposed for five minutes into the head space in order to sample the outgassing substances.

One sample of SP-50 ('04 configuration) has also been prepared in slightly different conditions. The aerogel tile has been cut into smaller fragments with respect to the previous experiments in order to sample a bigger amount (~2.4 grams) and the SPME fibre was exposed for 20 min. (Fig. 1b). Procedure blanks have been carried out as well.

After exposure, the SPME fibre was introduced into the injection port, maintained at ~ 250° C, of a gas chromatograph (Trace GC ultra, Finnigan-Thermo) coupled with an ion trap mass spectrometer (Polaris Q, Finnigan-Thermo). The MS ion source temperature was kept at 250°C. The mass spectrometer was operating in the EI positive mode (70 eV) with a mass range of 45-650 *m/z* (mass over electric charge). For the gas chromatographic separation a 5% diphenyl-95% dimethyl polysiloxane column (SPB-5, Supelco) (30m × 0.25mm i.d., 0.25 μm film thickness) was used in splitless mode. The chromatographic conditions were: 40°C (5 min.), 5°C/min., 250°C (20 min.). The carrier gas (He) was used in constant flow mode at 1.0 ml/min.

*2.2 Results*

In the chromatograms of all aerogel samples (Figs. 1-2 and Table 2) a low molecular weight siloxane has been revealed at RT 2.83. This compound is probably related to the chemical treatment of the aerogel used to confer hydrophobic properties.

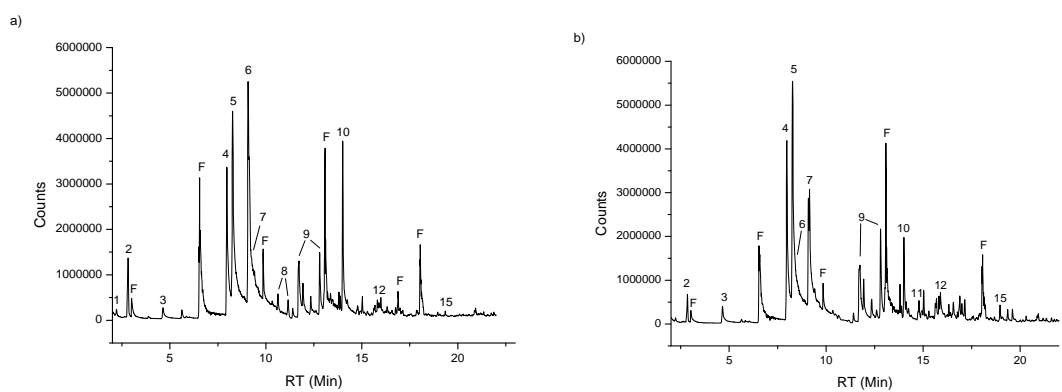

**Fig. 1: Chromatograms of Silicon aerogel samples. The two samples of SP50 a) unused ('05 configuration) and b) used ('04 configuration) have been analyzed with the same SPME-GC/MS technique but two slightly different procedures, read text for more details. The main identified compounds are listed in Table 2. Peaks of siloxanes, due to the SPME fibre, are labelled F in the chromatograms.**

**Table 2: Results of SPME-GC/MS analyses of silicon aerogel samples, see Fig. 1-2.**

| n. | Identified compound |
|---|---|
| 1 | Benzene |
| 2 | Siloxane |
| 3 | Toluene |
| 4 | o-Xylene |
| 5 | m,p-Xylene |
| 6 | Styrene |
| 7 | Ethylbenzene |
| 8 | Monoterpene |
| 9 | Ethylmethylbenzene |
| 10 | Monoterpene |
| 11 | Methylpropylbenzene |
| 12 | Ethyl-dimethylbenzene |
| 13 | Monoterpene |
| 14 | Diethyl-methylbenzene |
| 15 | Naphthalene |

The majority of the other peaks can be attributed to aromatic hydrocarbons; small amounts of benzene, toluene, naphthalene and various alkylbenzenes could be detected together with major amounts of xylenes, ethylbenzene and styrene. Terpenes (mono- and sesquiterpenes) were also revealed in relatively high amounts, especially in samples SP-15 and SP-30. All these compounds are "normally" present in ambient air. Styrene is also used in the manufacturing process of silicon aerogel.

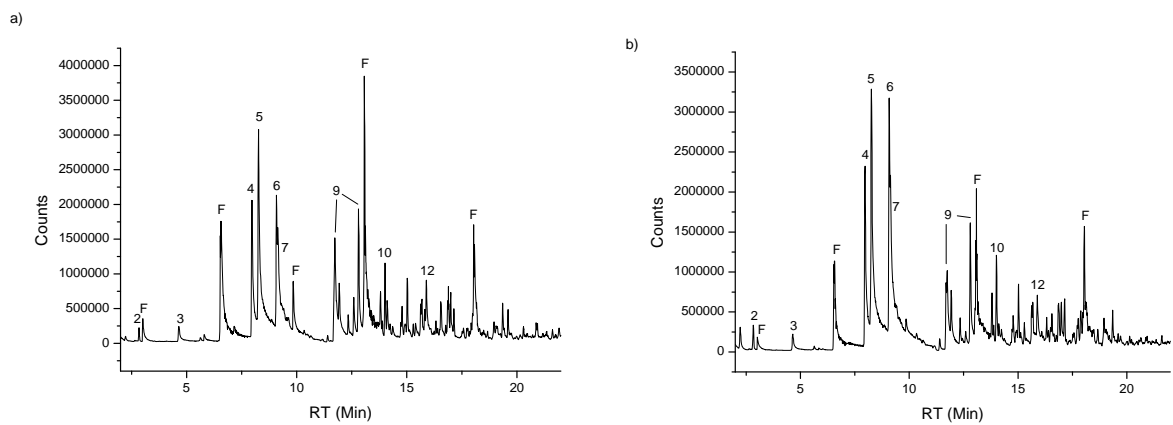

**Fig. 2: Chromatograms of Silicon aerogel samples. The samples of a) used ('04 configuration) SP-15 and b) unused ('05 configuration) SP30 have been analyzed with the same SPME-GC/MS technique and with the same procedure. The main identified compounds are listed in Table 2. Peaks of siloxanes, due to the SPME fibre, are labelled F in the chromatograms.**

Only slight differences could be observed between used ('04 configuration) and unused ('05 configuration) samples, except for the relative amounts of styrene with respect to xylenes. Styrene concentrations are higher in the unused samples. Normalised peak areas of styrene, ethylbenzene and xylenes against the peak area of a siloxane (RT 22.90), typical of the SPME fibre, have been calculated, see Table 3 and Fig. 3, in order to facilitate comparison between samples. Reproducibility of peak areas as obtained by SPME-GC-MS analysis is typically in the 10-15% range.

**Table 3: Results of SPME-GC/MS analyses of silicon aerogel samples. Peak areas of styrene, xylenes (o and m, p) and ethylbenzene have been normalised against the peak area of siloxane (RT 22.90) typical of the SPME fibre. Peak areas of styrene, xylenes and ethylbenzene have been calculated from single ion chromatograms (SIC) of *m/z* 104, *m/z* 106 and *m/z* 106 respectively, whereas siloxane areas have been determined from the total ion chromatogram (TIC).**

| Aerogel SP50 (Unused – '05 configuration) | siloxane | o-xylene | m,p-xylene | ethylbenzene | styrene |
|---|---|---|---|---|---|
| RT | 22,89 | 7,96 | 8,26 | 9,13 | 9,07 |
| Area | 3208226 | 1342713 | 3902625 | 825447 | 6372833 |
| Area norm | 1 | 0,42 | 1,22 | 0,26 | 1,99 |
| **Aerogel SP50 ('04 configuration)** | | | | | |
| RT | 22,91 | 7,95 | 8,26 | 9,13 | 9,07 |
| Area | 2385773 | 841597 | 1844706 | 524536 | 957403 |
| Area norm | 1 | 0,35 | 0,77 | 0,22 | 0,40 |
| **Aerogel SP15 ('04 configuration)** | | | | | |
| RT | 22,91 | 7,98 | 8,26 | 9,14 | 9,08 |
| Area | 4366142 | 1141315 | 2746656 | 719890 | 2434713 |
| Area norm | 1 | 0,26 | 0,63 | 0,16 | 0,56 |
| **Aerogel SP30 (unused – '05 configuration)** | | | | | |
| RT | 22,90 | 7,96 | 8,26 | 9,14 | 9,08 |
| Area | 3668949 | 1051104 | 2510635 | 584458 | 4054813 |
| Area norm | 1 | 0,29 | 0,68 | 0,16 | 1,11 |
| **Aerogel SP50 ('04 configuration) (Modified procedure)** | | | | | |
| RT | 22,90 | 7,97 | 8,27 | 9,14 | 9,08 |
| Area | 3459909 | 2276389 | 4828816 | 1257335 | 2788535 |
| Area norm | 1 | 0,66 | 1,40 | 0,36 | 0,81 |

The SPME-GC/MS analysis of the SP-50 sample prepared with a slightly modified procedure (increased sample quantity, smaller fragments, longer SPME fibre exposure time) generated a chromatogram with higher peaks but no qualitative differences.

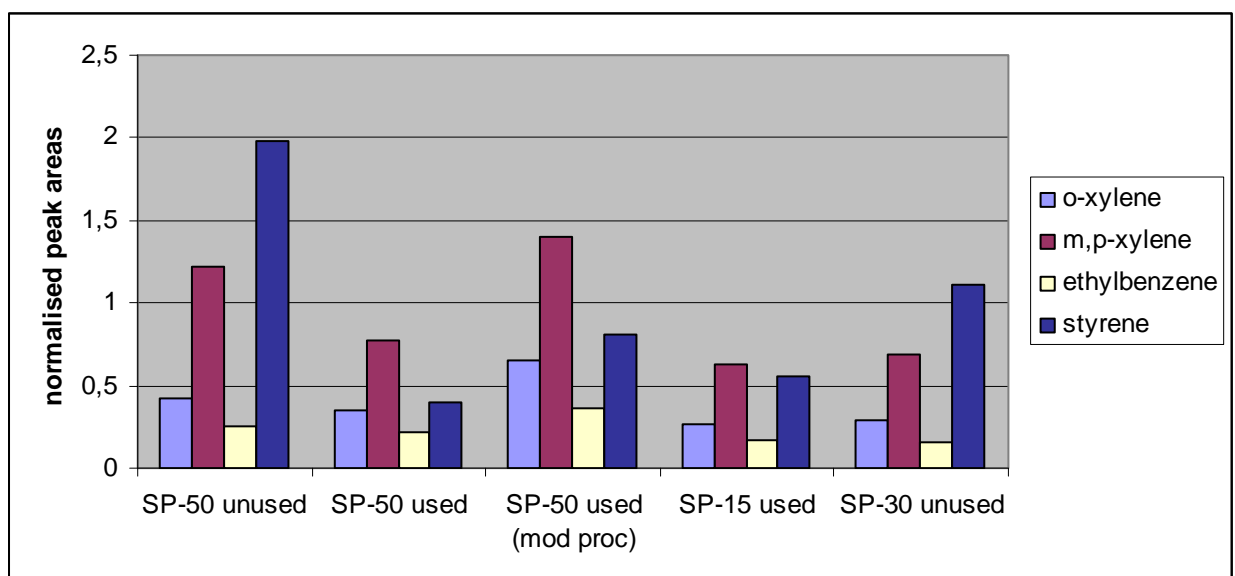

**Fig. 3: Results of SPME-GC/MS analyses of silicon aerogel samples. Peak areas (indicative of mass) of styrene and of xylene isomers (*o*, *m* and *p*) normalised against the peak area of a siloxane (RT 22.90 min.) typical of the SPME fibre used in this procedure. Normalisation is necessary in order to permit direct comparison between the different aerogel samples. The styrene/xylene ratio is inverted passing from non-contaminated (SP50('05) and SP30('05)) to contaminated (SP50('04) and SP15('04)) aerogels.**

The chromatogram of the adhesive tape, Table 4 and Fig. 4, shows that only few of the molecules detected in the aerogel samples are also present in smaller amounts in the adhesive tape. Vice versa the chromatogram of the adhesive tape shows some typical markers such as phenol and methylphenol that are completely absent in the aerogels.

**Table 4: Results of SPME-GC/MS analysis of a sample of adhesive tape as used in the setup of the '04 configuration (Fig. 4).**

| n. | Identified compound |
|---|---|
| 1 | Toluene |
| 2 | o-Xylene |
| 3 | m,p-Xylene |
| 4 | Ethylbenzene |
| 5 | Benzaldehyde |
| 6 | Phenol |
| 7 | Monoterpene |
| 8 | 2-hydroxybenzaldehyde |
| 9 | Methylphenol |
| 10 | Naphthalene |

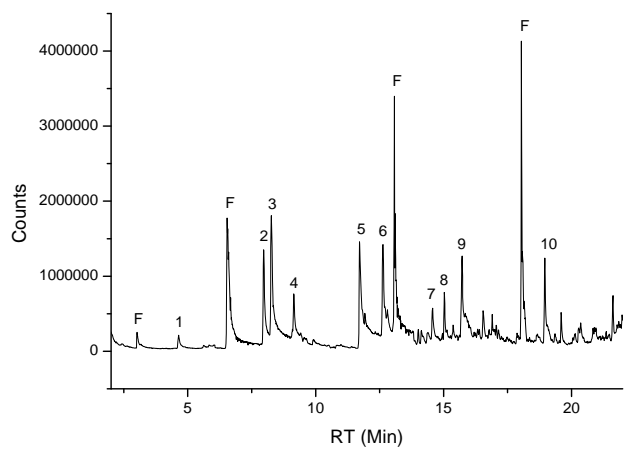

**Fig. 4: Chromatogram of adhesive tape as used in the setup of the '04 configuration. The main identified compounds are listed in Table 4. Peaks of siloxanes due to the SPME fibre are labelled F in the chromatogram.**

## 3. Contact Angle Analysis

In order to verify eventual differences in superficial hydrophobicity, measurements of contact angle have been performed by means of a contact angle goniometer (NRL model 100-00, Ramé-Hart). Two SP50 samples (used in '04 configuration and unused) have been analysed. For both samples a contact angle of 155° (average value of 5 measurements) has been determined and therefore no difference is detected.

## 4. Gas Chromatography of $CH_2Cl_2$ extract

*4.1 Experimental*

Samples of SP50 aerogel (used – '04 configuration - and unused – '05 configuration -), previously heated for SPME-GC/MS analyses, have been extracted with 20mL $CH_2Cl_2$ (20 min. ultrasonic bath). Aliquots of extract (4 mL) have been reduced to ~0.5 mL.

Amounts of 1 μL have been injected into the injector port, maintained at ~ 250° C, of a gas chromatograph (GC-17A, Shimadzu) with flame ionisation detector (FID). The chromatographic conditions were: 45°C (12 min.), 5°C/min., 220°C (20 min.). The carrier gas ($N_2$) was used with a linear flow rate of 15cm/s. A split ratio of 1:20 has been used.

*4.2 Results*

The GC analysis of the extracts of both samples of aerogel did not produce useful results. The chromatograms are similar to that of the solvent. No significant additional peaks have been detected.

## 5. UV-VIS spectroscopy

*5.1 Experimental*

Aerogel samples have been introduced into a UV-VIS spectrophotometer (UV-1601, Shimadzu). Transmittance spectra have been recorded from 220 to 800 nm. Measurements have been performed with the reference beam in air or by direct comparison between aerogel samples by inserting one of them as a reference. With the latter method small differences could be detected.

*5.2 Results*

Transmittance curves show a rapid decrease below 400 nm, confirming previous UV-VIS spectroscopy results. Direct comparison experiments revealed, however, major differences in transmittance behaviour in the 300-250 nm region, see Fig 5. It should be observed that aromatic hydrocarbons show maximum absorption peaks in the 300-250 nm region.

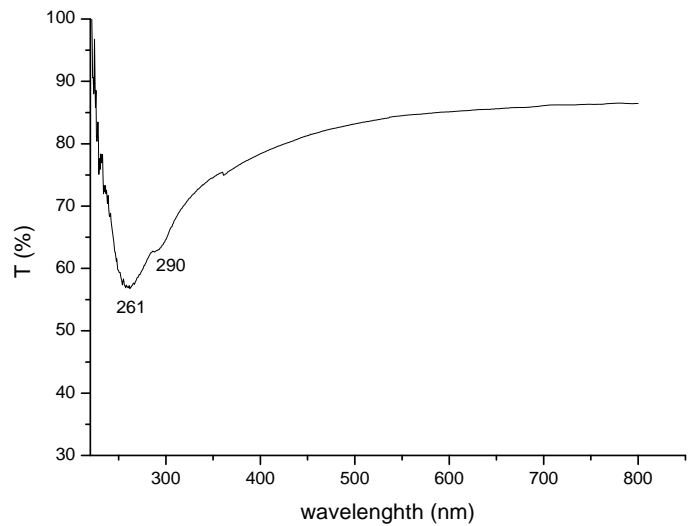

**Fig. 5: Transmittance curve as a function of the wavelength for the SP50 aerogel samples. The spectrum has been recorded by direct comparison of the used ('04 configuration) sample with the unused one (inserted in the reference beam).**

A systematic series of transmittance measurements on several Aerogel samples (old and new) have been performed in the D. Macchia Master Thesis [Mac05] using a Perkin-Elmer spectrophotometer. The optical quality of Aerogel samples is measured by the transmittance ($T$) curve, well accounted by the Hunt formula:

$$T = A \cdot \exp(-Ct/\lambda^4). \qquad (1)$$

*T* is the fraction of light as a function of the wavelength, *λ*, transmitted through a *t* thick tile. The Hunt parameters, *A* (transparency) and *C* (clarity, $\mu m^4$/cm units), account respectively for absorbed and Rayleigh scattered light [Asc2000]. Samples of high optical quality have *A* values very near to one and low values of *C*, near to zero. Fig. 6 illustrates the average ratio (Old/New) of those measurements in A1 and in A2. Also in this case it is evident that the higher absorption of the light starts around at 300 nm wavelength.

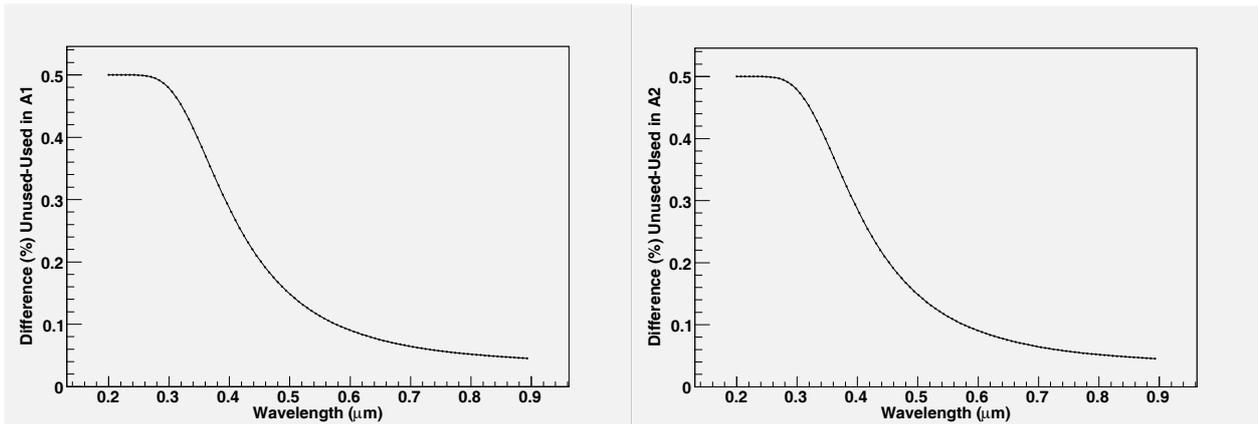

**Fig. 6: Differences in Transmission curves (unused-used in %) as a function of the wavelength for the SP15 aerogel samples (left graph) and SP50 (right graph). The difference is the average value between several measurements performed with the Perkin-Elmer spectrophotometer, read more details in Ref. [Mac2005].**

## 6. Conclusions

Finally in this section we will try to summarize the previous results providing some additional hints reported in other works to picture definitive conclusions.

The contaminated silica Aerogel is yellow because of its increased absorbance in the blue region with respect to the red part of the visible light, see Figs. 5 and 6. This effect is evident by naked eye when several layers of aerogel are stacked together while it is difficult to observe in a single tile.

The Matsushita aerogel is hydrophobic and it is demonstrated that this property is not lost because of ageing, see Ref. [Bel2003]. Moreover as illustrated in this work, see Sec. 3, the hydrophobicity over the tile surface is preserved after long time of detector operation. The irradiation of the tiles, due to the beam or to the sources, does not deteriorate the Aerogel optical qualities as reported in the references [Bel2003, Jac2005, Mar2005].

The Millipore paper and/or the adhesive tape cannot be the cause of the Aerogel contamination. In fact the presence of styrene and xylenes is probably not the cause of the light absorption because the amount of those molecules is sizeable in non-contaminated (unused) aerogel as well as in the old (used) one. In this case the air, flushed inside the counters, removes the molecules of styrene, used in the Aerogel manufacture, and introduces xylene molecules contained in the air. In fact the presence of styrene is more abundant in the new (unused) tile than in the old contaminated aerogel, see Figs. 1, 2 and 4 and Tables 2 and 3.

According to those evidences, the mentioned components: humidity, irradiation, tape, Millipore, cannot be the cause of the Aerogel contaminations.

The air, flushed inside the counter, is therefore the only cause of the Aerogel deterioration. In our measurements, there is a moderate transparency loss in the visible region and a strong light absorption in the blue-UV region, see Fig. 5 and 6. The same effects are detected in two other works, see Ref. [Gou1999, Mar2005], where the authors have contaminated the Aerogel by "aereosols fumes". In our case these "fumes" can be introduced in the counters just by the flushed air. The contaminants must have the size of sub-micrometric particles otherwise they would have been stopped by the filters or by the Aerogel itself (the pores of the Aerogel have diameter less than 100 nm). Moreover they are diffused in low concentrations in the whole detector, otherwise their presence should have been revealed by the chemical analysis.

A further confirmation of this picture is given by the performance of the RICH detector at HERMES experiment [Jac2005]. This detector has an Aerogel radiator of the same type (Matsushita SP30, $n$ =1.030) that has not deteriorated its optical properties in six years of operation [Del2007]. In that case pure nitrogen, instead of air, is flushed in the detector.